# Enhancement of hydrogen absorption and hypervalent metal hydride formation in lanthanum using cryogenic ball milling

Sakun Duwal[1], Vitalie Stavila[2], Catalin Spataru[2], Mohana Shivanna[2], Portia Allen[1], Timothy Elmslie[1], Christopher T. Seagle[1], Jason Jeffries[3], Nenad Velisavljevic[3], Jesse Smith[4], Paul Chow[4], Yuming Xiao[4], Maddury Somayazulu[4], Peter A. Sharma[1]

1. Sandia National Laboratories, Albuquerque, NM 87185,
2. Sandia National Laboratories, Livermore, CA 94550
3. Lawrence Livermore National Laboratories, Livermore, CA 94550
4. Argonne National Laboratories, Lemont, IL 60439

## Abstract

Rare earth superhydrides exhibit high temperature superconductivity but are difficult to characterize and use in applications due to their high formation and stability pressures, which are typically in excess of 100 GPa. We studied how modification of the rare earth precursor improves hydrogen reactivity and hydrogen uptake for forming such metal hydrides at lower pressures. An elemental lanthanum precursor was milled at liquid nitrogen temperatures for different time intervals. After exposure to gaseous hydrogen at 380 °C and 100 bar, we found a systematic enhancement of hydrogen absorption with increasing ball milling time for forming the $LaH_x$, x=2-3 phase. Exposing the precursor to pressures up to 60 GPa with an ammonia borane ($BNH_6$) hydrogen source resulted in a hypervalent $LaH_4$ phase. This $LaH_4$ phase is associated with the suppression of a rhombohedral distortion of the $Fm\bar{3}m$ cubic structure after cryomilling the precursor.

**Introduction**

Hypervalent metal hydrides, composed of rare earth or alkali metal species and large hydrogen-metal ratios, have been shown by experiments [1-7] and ab initio calculations [8-11] to support superconductivity at maximum transition temperatures of over 200 K. However, these materials are only stable at pressures in excess of 100 GPa, which greatly limits the available characterization techniques one can use to study this phenomenon. Applications are also not possible at these very high pressures. One would like to find a metal hydride that supports high temperature superconductivity that can be both synthesized and remain stable at far lower pressures, ideally at ambient pressure.

All of the rare earths normally form hydrides at ambient pressures with a stoichiometry of $REH_x$ with x up to ~3 with variable hydrogen/metal fraction limited by the number of available interstitials in the rare earth host lattice [12]. Solid hydrogen, which occurs at pressures greater than 600 GPa [13], has long been postulated to be a high temperature superconductor due to its light mass and potential for strong electron-phonon coupling [14]. The central idea behind realizing high temperature superconductivity in the metal hydrides is to synthesize a material with a covalently bonded hydrogen network supported by the metal sublattice [15]. For instance, the hydrogen sublattices in these materials often take on exotic structural motifs such as cages [16]. Thus, large amounts of hydrogen are needed in the "supervalent" metal hydrides and the metal sublattice lattice must expand in volume to accommodate larger hydrogen/metal fractions to potentially realize novel hydrogen bonding states. These supervalent rare earth hydrides require such large pressures due to the combination of kinetic barriers to insert hydrogen and the thermodynamic stability of the crystal structures.

In order to address the problem of achieving supervalent hydrides at lower pressures, prior research has focused on computing the transition temperature and formation pressure for a wide variety of different binary and ternary hydrides using *ab initio* methods to inform further experimentation of equilibrium compounds [17-19]. Doping of an existing hydride is also being explored on both the rare earth and hydrogen sites for this purpose [20-23]. Both of these prior approaches have shown how phase stability of supervalent superconducting hydrides can be achieved at lower pressures. In this work, we explored an alternate possible route for forming these materials at lower pressures through a processing technique, mechanical milling, employed on an existing material.

Milling is one of many processing methods used to achieve phase transformations that are far out of equilibrium under prevailing conditions [24, 25]. High pressure phases of binary rare earth alloys [26] and elemental lanthanum [27] have been stabilized under ambient conditions using mechanical milling. Milling techniques are also commonly used to form non-equilibrium phases that enable higher hydrogen loading in metal hydrides for hydrogen storage [28, 29]. The reasoning behind mechanical milling is motivated by the large pressures and temperatures experienced between two colliding spherical particles with a small contact area [30], which can be viewed as an alternative way of exposing a material to a (transient) high pressure environment followed by a rapid quench. The purpose of the present work was to understand if a non-equilibrium structural phase of a supervalent hydride could be promoted using a mechanical milling process in analogy with the empirical metallurgical observations in intermetallics and metal hydride materials.

We performed cryogenic ball milling of an elemental lanthanum over different durations and then exposed this modified precursor to hydrogen. Cryogenic ball milling was required due to the ductility of the rare earths at room temperature. After forming cryomilled materials, we studied the hydrogen absorption properties of the resulting hydride at pressures of 100 bar and a temperature of 380 °C. Hydrogen diffused faster into the ball milled materials and increased the hydrogen loading relative to the as received powder. We then studied the formation of any higher hydrides up to 60 GPa with synchrotron X-ray diffraction using diamond anvil cells and found a hydrogen deficient $LaH_4$ phase with an expanded unit cell, formed at room temperature. Cryogenically ball milled precursors increased the pressure at which we observed this distorted unit cell but increased the hydrogen content in the supervalent metal hydride. We thus formed a non-equilibrium supervalent metal hydride by modifying the precursor rare earth using mechanical milling followed by exposure to hydrogen at high pressures. This result thus shows that suitable modification of the metal precursor can change the thermodynamic stability for a supervalent hydride.

**Methods**

La powder was purchased from American elements with a purity of 99.9% and mesh size 325. As received La fine black color powder of approximately 1 g was transferred to a milling vessel with the addition of five 5 mm steel balls. The powder entirely covers the balls in order to achieve efficient mixing. The milling vessel was exposed to a continuous liquid nitrogen feed in order to maintain cryogenic temperatures. Milling of the precursor was performed in a glove box under an argon environment.

Powder X-ray diffraction (PXRD) was performed in a sealed 0.6 mm capillary. PXRD experiments were carried out on a PANalytical Empyrean™ diffractometer equipped with a PIXcel3D detector and Cu K-alpha radiation ($\lambda\alpha$ = 1.5418 Å). The diffractometer was operated at 44 kV and 40 mA in continuous scanning mode with the goniometer in the theta-theta orientation.

Particle size measurements were performed using dynamic light scattering in a Horiba Partica LA-960V2. Cryomilled powder was ultrasonicated in 200 proof ethanol. Three repeated measurements were taken for each sample.

High pressure synchrotron X-ray diffraction experiments were performed at the Advanced Photon Source beamlines 16-ID-D and 16-ID-B. We used a symmetric diamond anvil cell (DAC) with flat culets of 300 μm diameter and a 140 μm diameter sample chamber drilled in a rhenium gasket for all experiments. Ammonia borane ($BNH_6$) was used as an internal hydrogen source where indicated, which irreversibly decomposes to BN and $H_2$ gas. While full decomposition is known to occur at 500 C [31], we found that at ambient conditions, there was sufficient hydrogen gas evolved to form hydrides readily within days. Ammonia borane/metal ratios were prepared in a 10:1 ratio by mass. The molar mass of ammonia borane is 30.865 g/mole, while that of La is 138.9 g/mole. Mixing the ammonia borane and lanthanum in a 10:1 ratio would then result in more than 24x more hydrogen than lanthanum atoms, assuming full decomposition of $BNH_6$ and no extraneous loss of hydrogen during packing and loading. We observed $LaH_{2+x}$ to form readily when mixed with this compound under ambient pressure and temperature conditions. All DAC samples were loaded and sealed within a glove box in argon atmosphere to avoid any exposure to oxygen and water. Further experimental details are listed in the Supplementary Information.

We carried out DFT calculations of several lanthanum hydride structures ($LaH_x$ with x=1,2,3 and 4) using the VASP code [32] in conjunction with projector augmented wave pseudo-potentials [33]. The exchange and correlation terms were treated within the generalized gradient approximation (GGA) with the Perdew-Wang parameterization [34] and the energy cutoff was set to 520 eV. Enthalpies for different structures considered were obtained by relaxing the forces (at fixed pressure) to better than 0.01 eV/Ang, using primitive unit cells with the Brillouin zone sampled with a 6x6x6 Monkhorst-Pack k-point grid.

**Results**

The primary control variable in this study was the amount of time spent milling the starting lanthanum material at liquid nitrogen temperatures. In order to understand if there were any changes in the material with respect to hydrogen interactions as a result of this process, we performed hydrogen absorption measurements using the Sieverts technique at a temperature of 380 C and a pressure of 100 bar in pure $H_2$ gas, as shown in Fig. 1. These absorption measurements were carried out for a maximum time of 600 minutes in a fixed volume under pure (99.99%) $H_2$ gas in order to ensure equilibrium and allow a direct characterization of the amount of hydrogen absorbed in a metal precursor, as described elsewhere. Figure 1 shows that both the rate of hydrogen absorption and the total amount of hydrogen in the lanthanum precursor (inset) increase with increasing milling time. The as received crystal structure of lanthanum was face centered cubic and did not change for any cryomilling or post $H_2$ annealing condition as determined from ambient pressure X-ray diffraction measurements. Dynamic light scattering of cryomilled powder showed a systematic decrease in median particle size of 34 μm, 10 μm, 2 μm for as received, 120 minutes, and 300 minutes, respectively. Precursor cryomilled powders generally had minimum particle sizes of ~500 nm. Representative data are shown in the Supplementary Information. For the rest of the paper, we focus on the changes in hydride formation at high pressures.

Figure 2 shows the evolution of the crystal structure of lanthanum hydrides under ammonia borane exposure formed from cryomilled La precursor. Cryomilled lanthanum was exposed to pressure up to 60 GPa with solid ammonia borane powder as a source of hydrogen, mixed by mortar and pestle in a 1:10 ratio (La:$BNH_6$) in an inert environment. X-ray diffraction was recorded at room temperature as the pressure was increased incrementally and then upon decompression at the Advanced Light Source, Argonne National Laboratory (HPCAT). Exposure of all La precursors to ammonia borane resulted in the $Fm\bar{3}m$ $LaH_x$ phase, indicating rapid hydrogenation at room temperature. The lattice parameter of this starting $LaH_x$ phase was ~5.648 Å, irrespective of cryomilling condition, pointing to a starting hydrogen composition of $LaH_{2.2}$ [35]. We observed all hydride samples to phase separate into compounds with higher and lower hydrogen content [36]. For clarity, we only show the higher hydrides in Fig. 2. More complete details, including example X-ray diffraction data and the different majority and minority phases, are included in the Supplementary Information.

Upon compression, this $Fm\bar{3}m$ structure transformed to a $R\bar{3}m$ structure (left panel). The formation of the $R\bar{3}m$ structure was accompanied by the coexistence of a lower $LaH_x$ phase, x~1, described in the Supplementary Information. We plot the lattice parameters of the $R\bar{3}m$ structure in the hexagonal setting, which defines an $a_H$ and $c_H$ lattice parameter. The pressure at which this phase transformation occurred increased with longer precursor cryomilling times. The body-diagonal distortion of the cubic $Fm\bar{3}m$ structure in this $R\bar{3}m$ phase can be quantified through the c/a ratio (right panel). For a and c lattice parameters in the hexagonal setting, the ideal $Fm\bar{3}m$ structure has a c/a ratio of 2.45 by definition. The as received La precursor, exposed to ammonia borane under pressure, first distorts to c/a ~2.65 then decreases to ~2.55. In contrast, the cryomilled samples maintain c/a ~2.65 up to 60 GPa. The increase in the critical pressure at which the distorted fcc phase was observed indicates that cryomilling aided in stabilizing the undistorted fcc phase at higher pressures compared to the control sample.

Figure 3 compares the cell volume per formula unit as a function of pressure of our cryomilled material relative to the as received La control sample. The cell volume of each material was extracted from Fig. 2 and normalized by the number of atoms in the unit cell. We performed reference experiments on pure lanthanum (solid black squares) and checked that the results agree with the prevailing literature [37, 38]. Our data for LaH$_{2.2}$ up to 20 GPa agrees with prior literature on a hydride with very similar composition [36]. The cell volume per lanthanum increases with longer cryomilling times (blue and red open squares for 60 and 120 minutes, respectively) compared to the as received control. The stoichiometry of this $R\bar{3}m$ phase at high pressure was inferred to be LaH$_x$ with x~4 based on the excess cell volume compared to pure La, as outlined in the Supplementary Information.

To further understand the thermodynamic stability of this structure given an inferred hydrogen content, we performed *ab initio* calculations of different LaH$_x$ compounds with the observed structural evolution. In order to obtain the hydrogen positions, we optimized the structure within DFT at fixed pressures using the stoichiometry LaH$_4$. We ignored the fact that the hydrogen stoichiometry was variable, in particular that the hydrogen/metal ratio only reaches ~4 at 60 GPa, since at room temperature hydrogen positions are likely statistically random over possible lattice sites and the symmetry of the La atoms does not change from x~3-4 in our data. Zero-point effects of hydrogen were also discounted. From these optimized structures, we then computed the equation of state for each of the phases participating in the reaction, which have been overlayed with the experimental data in Figure 3 over the entire pressure range. The equation of state of the 120 minute cryomilling sample more closely matches the $R\bar{3}m$ prediction for LaH$_4$. The XRD data thus point to an LaH$_4$ phase formed by cryomilling at high pressures based on the excess cell volume and DFT equation of state calculations. We also found a $R\bar{3}m$ structure of LaH$_3$, but the equation of state and enthalpy for this phase were nearly identical to the $Fm\bar{3}m$ structure as the rhombohedral distortion was very small. Values of x>3 in LaH$_x$ are supervalent in the sense that the hydrogen content exceeds the maximum possible hydrogen/metal fraction for an fcc structure of the host La atoms in the LaH$_3$ phase. The as received and 60 minute samples also show a distorted R$\bar{3}$m structure, so the enhanced hydrogen absorption in the 120 minute sample is not due to the structural change alone. The structural changes observed in Fig. 2 and 3 were irreversible upon decompression.

We studied the stability of the observed LaH$_4$ phase relative to cubic LaH$_3$ by computing the enthalpy of possible reactions amongst hydride phases as a function of pressure We assumed the reaction $LaH_2 \rightarrow \alpha LaH + \beta LaH_x$, and fixed the volume fraction of the solid solution phase ($\alpha$) and the higher hydride ($\beta$) in order to conserve hydrogen stoichiometry. The stoichiometry x was fixed at 3 and 4 to match the observed high pressure (~60 GPa) value. The enthalpy of each possible reaction was then computed and plotted as a difference of the products and LaH$_2$ in Fig. 4. This plot shows that the decomposition of LaH$_2$ to LaH$_3$ and LaH becomes exothermic at approximately 10 GPa. This result agrees with prior experimental and DFT studies [39], and thus benchmarks our calculations. The present calculations show that the LaH$_4$ phase becomes exothermic relative to LaH$_3$ at higher pressures of ~30 GPa. However, the normative LaH$_3$ phase was calculated to be stable over the entire pressure range studied here (up to 60 GPa). Thus, these DFT calculations show that the La sublattice structure we observe with a hydrogen/metal ratio greater than 3 does not appear to be an equilibrium phase.

**Discussion**

Mechanical ball milling is a process whereby a starting precursor is subjected to high energy impacts in order to accomplish either comminution (reducing particles sizes) or attrition (plastic deformation and damage to a particle of a given size) [40]. A powder charge is added into a sealed metal chamber with spherical milling balls, which then undergoes vibrations in order to generate collisions. Cryogenic temperatures are used to avoid local heating and recrystallization and/or to make the milled material more brittle in soft materials such as plastics or soft metals, enabling fracture [41]. In the present context, a high transient pressure can be realized from the collision of two spherical objects if they each have a sufficiently high bulk modulus [30]. Our SEM observations (Supplementary Information) indicate some reduction and homogenization of particle size, but comminution and attrition likely both occur.

The ball milling process increased the kinetics and thermodynamic stability for larger hydrogen loading of the lower lanthanum hydride phase as shown in Fig. 1. The kinetics of hydrogen reaction with a metal involves several steps. Di-hydrogen gas must first be transported to the surfaces, followed by dissociation into atomic hydrogen (possibly at special sites on the surface). This atomic hydrogen must then penetrate any surface barriers, such as oxides, and diffuse into the metal to form a surface hydride layer. Further diffusion must then occur through any metal hydride surface layers. For example, on oxidized surfaces, $H_2$ dissociation and subsequent permeation through the oxide greatly limits the kinetics [42]. Once through the oxide layer, hydrogen usually diffuses rapidly through the metal. The diffusivity of hydrogen in $LaH_x$ for x~2.5 is approximately $1 \times 10^{-10}$ $m^2/s$ at 380 C [43], corresponding to diffusion over 100 microns in 1 ms, and approximately an order of magnitude greater along grain boundaries [44].

The data in Fig. 1 suggest our starting powder is surface limited. The ball milling process increased the absorption rate of hydrogen with increased milling time, which could be due to an increased surface area and/or activation site density for hydrogen dissociation. In addition to an increase in hydrogen absorption rate, the amount of hydrogen absorbed at saturation also increased with increasing milling time. The low pressure La-H system has three main phases [12, 45]. At low hydrogen-metal fractions of <5 at. %, hydrogen is distributed randomly over the tetrahedrally coordinated sites of the fcc La lattice. Increasing the hydrogen content results in a eutectic phase diagram between this hydrogen-doped La phase and a non-stoichiometric homogeneous hydride $LaH_x$ for x~1.9-3.0 The $LaH_2$ phase has the flourite fcc structure with hydrogen occupying a large, variable fraction of the tetrahedral sites. Further hydrogenation occurs until $LaH_3$ is formed, the maximum for the fluorite structure where both the octahedral and tetrahedral interstitial sites are fully occupied by hydrogen. The trend towards higher hydrogen fractions of $LaH_3$ in Figure 1 shows that the ball milling process has increased the occupation of the more inaccessible octahedral sites.

Our primary observation is the realization of a higher lanthanum hydride phase promoted through the application of mechanical ball milling of the lanthanum precursor. This higher hydride phase was realized at room temperature under pressure. Cryomilling delayed the onset of the distorted fcc phase as shown in Fig. 2, highlighting that metastability has been introduced. The stoichiometry of this phase (Fig. 2 and 3) is consistent with a hydrogen content x>3 for $LaH_x$ based on the volume of the unit cell per formula unit and the distorted fcc structure inferred from the X-ray diffraction data. Hydrogen/metal ratios greater than 3 are considered supervalent here in that the hydrogen must formally exist in one or more combinations of bonding states (interstitial H0, H-, H+, or covalently bound H). For example, intermetallic hydrides with an anisotropic structural distortion, such as that observed in this work, show a mixed ionic/covalent hydrogen bonding character [46]. The maximum hydrogen/metal ratio, x, observed here is ~4 at the highest pressures studied (60 GPa). However, the data show x >3 at lower pressures and structural changes were irreversible upon decompression. Recognizing that hydrides have significant variability in hydrogen content, we assign this phase as LaH$_4$. DFT calculations of this LaH$_4$ phase show an equation of state consistent with the data in Fig. 2.

Various alkali tetrahydrides exist [47, 48] with the I4/mmm structure below 60 GPa, the maximum pressure for measurements reported here. Lanthanum tetrahydrides were predicted to have a tetragonal I4/mmm structure [8, 49], confirmed experimentally at pressures of >100 GPa [50]. We indexed our XRD data to a rhombohedral structure, i.e. judging from the La positions alone. While we can only speculate on hydrogen positions from the optimized DFT prediction with fixed La, the distorted fcc lattice for the structure we have observed is not close packed and thus can accommodate more than the maximum 3:1 ratio of interstitial to atomic sites expected for an fcc structure. There were no previous literature reports on the behavior of LaH$_3$ in the pressure range of 30-60 GPa. While further study of this LaH$_4$ phase is required to identify the structure more accurately, cryomilling the La precursor appears to promote the formation of this phase at lower pressures and furthermore this phase must be metastable at pressures <60 GPa judging from our control experiments and DFT calculations. Electronic structure calculations (Supplementary Information) show that this supervalent phase has a finite density of states at the Fermi level with a contribution from both La and H orbitals and thus it may support hydrogen-promoted superconductivity.

**Conclusion**

We studied how milling a lanthanum precursor altered the hydrogen absorption properties and formation of supervalent hydrides at high pressures. Cryomilling systematically enhanced the hydrogen absorption kinetics and hydrogen/metal fraction at 380 C and 100 bar of H$_2$ gas. Cryomilling-induced metastability was also observed in the high pressure phase transformations of LaHx powder at room temperature. The formation of a supervalent hydride, LaH$_4$, with a rhombohedral phase was observed at >30 GPa under hydrogen exposure using ammonia borane through the application of cryogenic mechanical ball milling of La powder at a maximum milling time of 120 minutes. DFT calculations were used to propose optimized hydrogen positions and support identification of the supervalent phase through equation of state modeling. Prior synthesis of supervalent hydrides require much higher temperatures and/or pressures than what has been achieved with this cryomilling processing route. We expect this processing route to be useful for any supervalent metal hydride. Modification of a metal precursor in general may be a useful route to lower the formation pressure and promote metastability at lower pressures in supervalent hydrides.

**Acknowledgements**

Sandia National Laboratories is a multi-mission laboratory managed and operated by National Technology & Engineering Solutions of Sandia, LLC (NTESS), a wholly owned subsidiary of Honeywell International Inc., for the U.S. Department of Energy's National Nuclear Security Administration (DOE/NNSA) under contract DE-NA0003525. This written work is authored by an employee of NTESS. The employee, not NTESS, owns the right, title and interest in and to the written work and is responsible for its contents. Any subjective views or opinions that might be expressed in the written work do not necessarily represent the views of the U.S. Government. The publisher acknowledges that the U.S. Government retains a non-exclusive, paid-up, irrevocable, world-wide license to publish or reproduce the published form of this written work or allow others to do so, for U.S. Government purposes. The DOE will provide public access to results of federally sponsored research in accordance with the DOE Public Access Plan. This research used resources of the Advanced Photon Source, a DOE Office of Science User Facility operated for the DOE

Office of Science by Argonne National Laboratory under contract no. DE-AC02-06CH11357. The instrumentation and facilities used were supported by DOE/BES (DE-FG02-99ER45775, VVS), the U.S. DOE/National Nuclear Security Administration (HPCAT). Work was also performed under the auspices of the US DOE by Lawrence Livermore National Laboratory under contract DE-AC52-07NA27344. We acknowledge M. Blea and Y. Meng for experimental support.

Figures

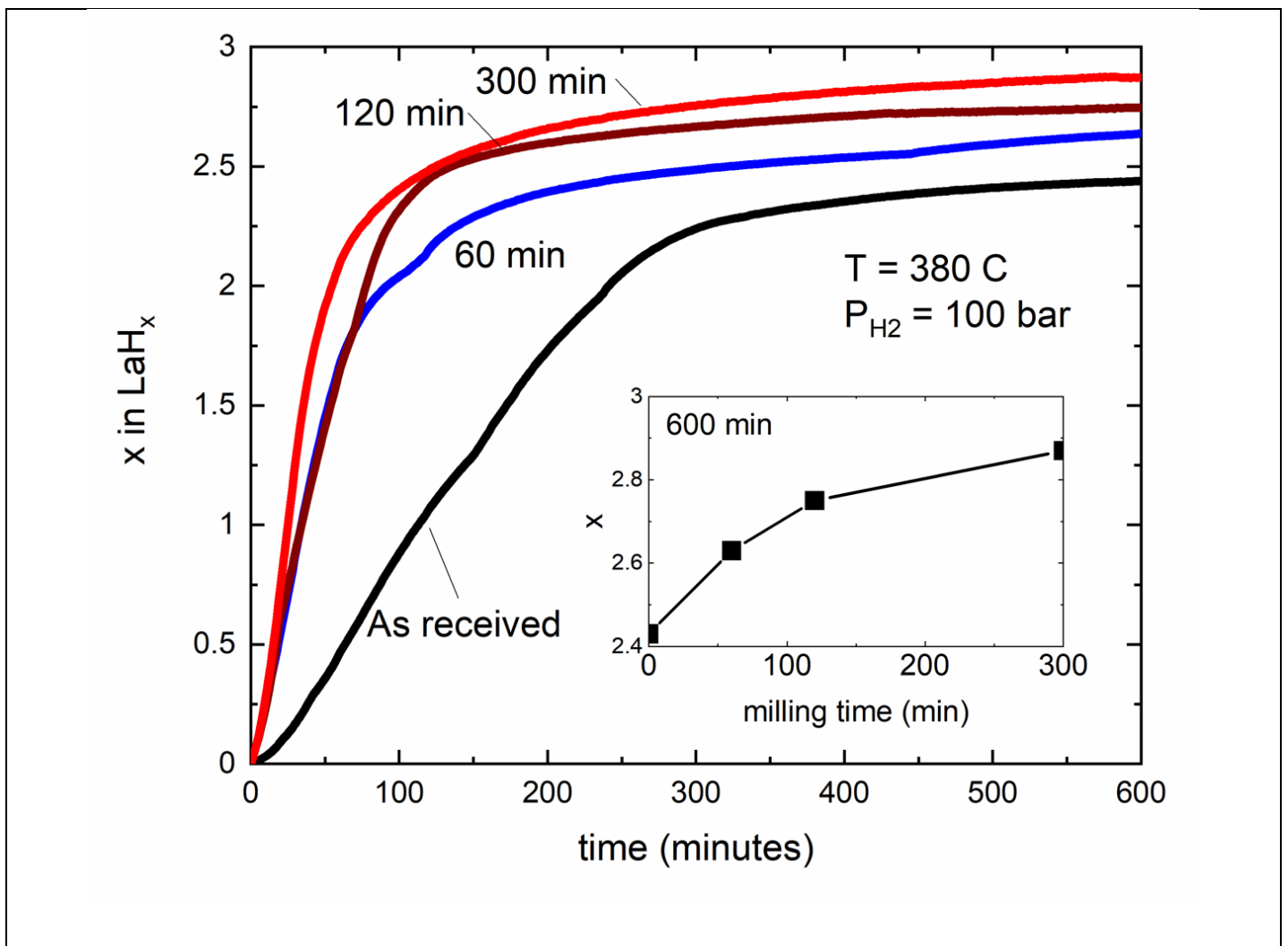

**Figure 1**. Isothermal hydrogen absorption as a function of time at a fixed $H_2$ pressure of 100 bar with T = 380 C for La powder cryomilled at different lengths of time. Compared to the as received La powder, cryomilled precursor materials systematically absorb hydrogen at a faster rate and absorb a larger total amount of hydrogen (inset).

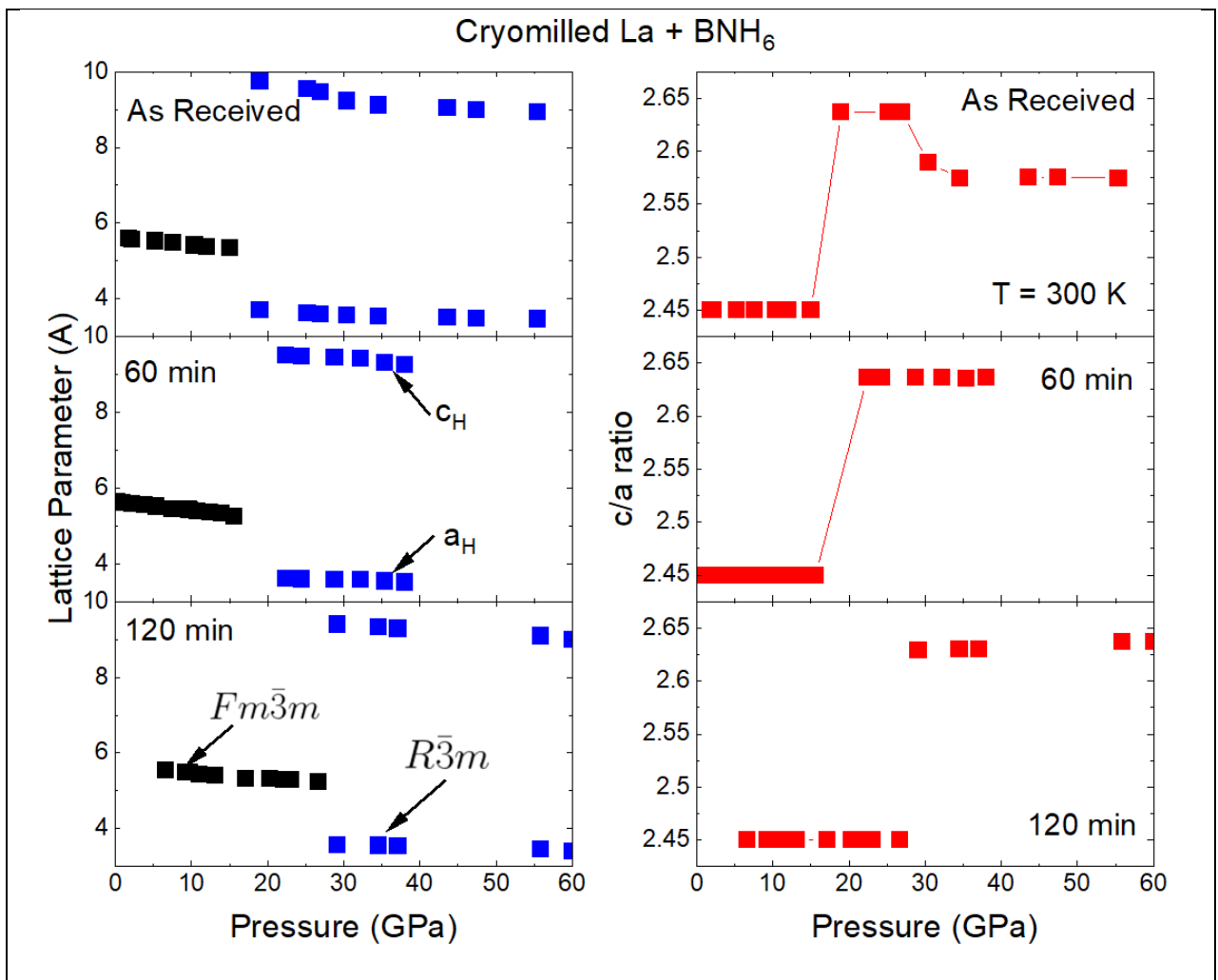

Figure 2. Lattice parameters (left) and c/a ratio for cryomilled La mixed with ammonia borane at room temperature. The hydride formed from this La material in the presence of $H_6BN$ undergoes a distortion from $Fm\bar{3}m$ to $R\bar{3}m$. Cryomilling the La precursor changes the pressure where this phase transformation occurs. The a and c parameters are defined in the hexagonal setting of the $R\bar{3}m$ phase. The c/a ratio is highest for the cryomilled samples compared to the as received a precursor. The c/a ratio in the ideal $Fm\bar{3}m$ structure is 2.45 by definition.

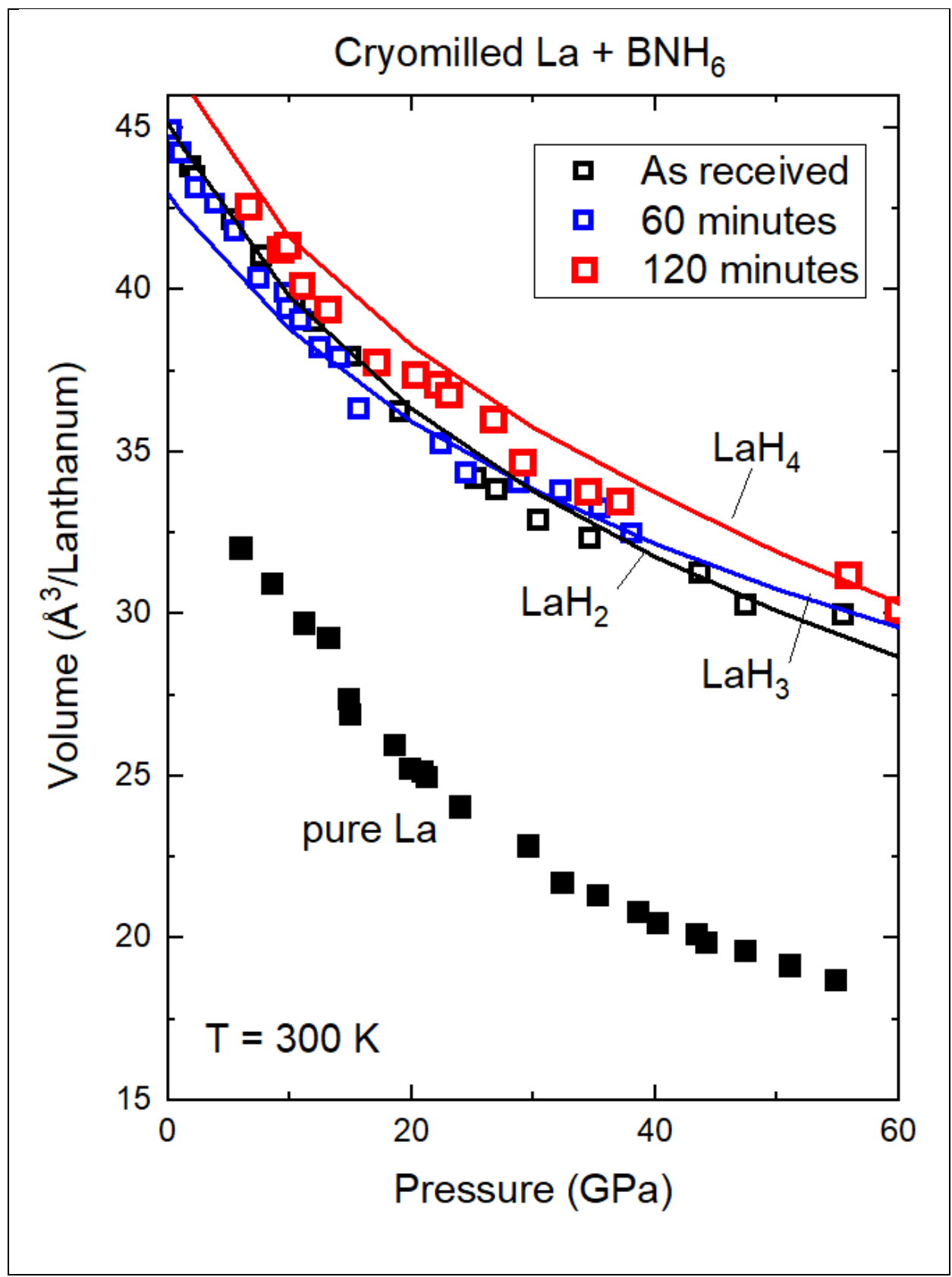

**Figure 3.** Equation of state (P-V) measurements for cryomilled La mixed with ammonia borane at room temperature. The unit cell volume was extracted from X-ray diffraction data for cryomilled samples (open red and blue squares) and compared with as received La (open black squares) when mixed with ammonia borane. The as received La was measured without ammonia borane (closed squares) and agrees with the prevailing literature. The equation of state was calculated using DFT for the known and proposed structures for $LaH_2$, $LaH_3$, and $LaH_4$ (black, blue, and red lines) and compared with the data.

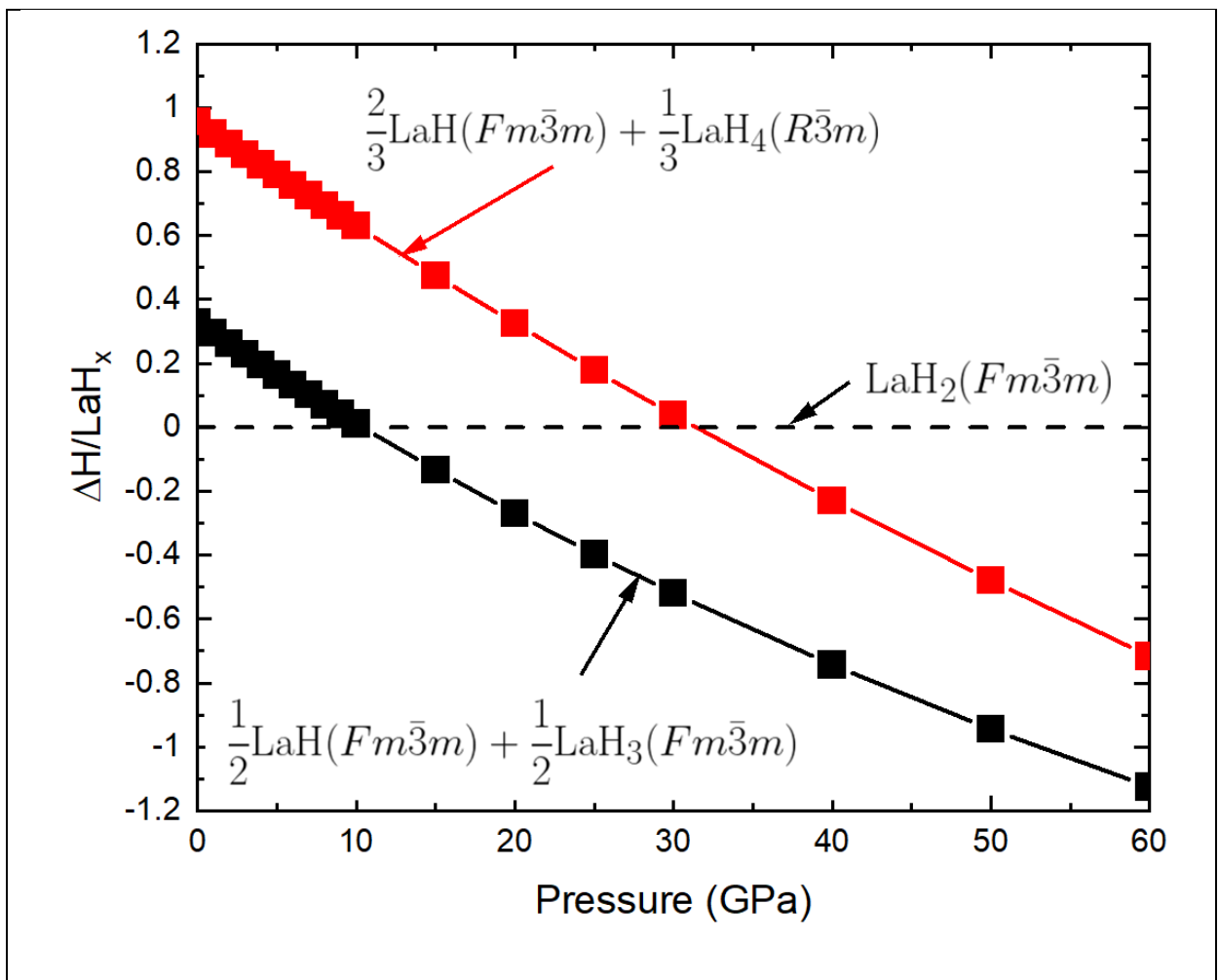

Fig. 4. Ground state enthalpy difference per formula unit relative to the Fm3-m LaH$_2$ structure over the entire pressure range studied. Decomposition of the lower hydride LaH$_2$ into LaH and LaH$_3$ is energetically favorable over decomposition into LaH and LaH$_4$, where the LaH$_4$ phase has a R$\bar{3}$m structure. Calculations for the other phases assume Fm$\bar{3}$m. We observe a critical transition of ~30 GPa for phase separation in Fig. 2.